\documentstyle[preprint,aps,eqsecnum,epsf]{revtex}

\def \en{\end{eqnarray}}
\def \bg{\begin{eqnarray}}
\def \enm{\end{mathletters}}
\def \bgm{\begin{mathletters}}

\def \r{\rho}

\def \Orc2{{1\over c^2}}

\def \Oc2{O(1/c^2)}

\begin{document}

\centerline{\bf TWO-BODY DIRAC EQUATIONS FOR} \centerline{\bf RELATIVISTIC
BOUND STATES OF QUANTUM FIELD THEORY} \centerline  {Horace W. Crater} %
\centerline{The University of Tennessee Space Institute} %
\centerline{Tullahoma,Tennessee 37388 } \centerline  {Peter Van Alstine} %
\centerline{12474 Sunny Glen Drive, Moorpark, Ca. 93021}
\begin{abstract}
\noindent

We review a little-known treatment of the relativistic two-body bound-state
problem - that provided by Two-Body Dirac Equations obtained from constraint
dynamics. We describe some of its more important results, its relation to
older formulations and to quantum field theory. We list a number of features
crucial for the success of such a formulation, many of which are missing
from other methods; we show how the treatment provided by Two-Body Dirac
Equations encompasses each of them.
\end{abstract}

\vspace{2cm}

\vskip 1 truein Invited paper presented at a Conference on September 12th,
1997 at the University of Georgia in honor of Professor Donald Robson on his
60th birthday. 

\vfill\eject
\newpage

\section{Introduction}

\setcounter{section}{1}

The correct formulation of relativistic two-body bound state wave equations
and their connection to quantum field theory is an old problem going back to
papers by Eddington and Gaunt \cite{edgnt} in 1928 (bases on conjectures by
Heisenberg). But judging from the large variety of approaches \cite{list}
attempted even in recent years, this problem has no generally agreed-upon
solution. Perhaps for this reason, recent field theory books have avoided
this topic. For example, Steven Weinberg \cite{wnbrg} states: ``It must be
said that the theory of relativistic effects and radiative corrections in
bound states is not yet in satisfactory shape.''

In \cite{list} we have listed many of the attempts at solving this problem
made over the past 47 years. At the head of this list, appear the work of
Nambu and the better-known work of Bethe and Salpeter, both developed over
20 years after the partly successful, semirelativistic equation of Breit.
The Bethe-Salpeter equation is an integral equation in momentum space that
is manifestly covariant, obtained directly from relativistic quantum field
theory. Over the years, however, many problems have turned up to impede its
direct implementation. These are the sources of the numerous attempts to
reformulate the two body problem of relativistic quantum field theory. For
example as pointed out by Weinberg \cite{wnbrg} : ``The uncrossed ladders
can be summed by solving an integral equation known as the Bethe-Salpeter
equation, but there is no rationale for selecting out this subset of
diagrams unless both particles are non-relativistic, in which case the
Bethe-Salpeter equation reduces to the ordinary nonrelativistic
Schr\"{o}dinger equation, plus relativistic corrections associated with the
spin-orbit couplings that can be treated as small perturbations.''
Furthermore, the Bethe-Salpeter equation in the ladder approximation
possesses negative norm or ghost states, due to its treatment of the
relative time degree of freedom - spoiling the naive interpretation of it as
a quantum wave equation.

Salpeter and many others \cite{list} have developed noncovariant
instantaneous truncations of this equation. For a variety of reasons most of
the attempts we have cited are not appropriate for the treatment of highly
relativistic effects like those necessary for the calculation of
quark-antiquark bound states.

Among the authors who have tried to rectify this problem are Professor
Robson and his collaborator Dr. Stanley who gave one of the first
comprehensive attempts to obtain a potential-model description of the entire
meson spectrum, combining exact relativistic kinematics with a
non-perturbative treatment of the effects of the important spin-dependent
interactions. On the other hand, our work treats much the same spectral
phenomenology but remains close to Dirac's own one body work by starting
with a pair of covariant two-body Dirac equations (one for each particle)
which forces certain restrictions on the various spin-dependent interactions
that can appear.

\section{Two-Body Dirac Equations from Constraint Dynamics}

The list of references on the relativistic two-body problem includes diverse
alternative approaches. On the other hand the one-body Dirac equation \cite
{Dirac} has no serious rivals. That equation is a well defined wave equation
that can be solved nonperturbatively and serves as an example of a
successful bound state equation.

\subsection{The One-Body Dirac Equation}

The free Dirac equation 
\begin{eqnarray}
(\gamma \cdot p+m)\psi =0
\end{eqnarray}
provides a relativistic version of Newton's first law, with associated
relativistic second law appearing as the four-vector substitution 
\begin{eqnarray}
p_\mu \to p_\mu -A_\mu
\end{eqnarray}
for electromagnetic interaction and as the minimal mass substitution 
\begin{eqnarray}
m\to m+S
\end{eqnarray}
for scalar interactions combining to give 
\begin{eqnarray}
\label{obd}
(\gamma \cdot (p-A)+m+S)\psi =0.\ref{obd}
\end{eqnarray}
As we shall see, the two-body Dirac equations take simple forms that
generalize this one to the interacting two-body system.

\subsection{The Breit and Eddington-Gaunt Equations}

The Breit equation \cite{Breit} and most three dimensional truncations of
the Bethe-Salpeter equation are neither well defined wave equations. nor
manifestly covariant. Breit proposed his equation equation

\begin{eqnarray}
E\Psi =\left\{ \vec{\alpha}_1\cdot \vec{p}_1+\beta _1m_1+\vec{\alpha}_2\cdot 
\vec{p}_2+\beta _2m_2-\frac \alpha r[1-\frac 12(\vec{\alpha}_1\cdot \vec{%
\alpha}_2+\vec{\alpha}_1\cdot \hat{r}\;\vec{\alpha}_2\cdot \hat{r})]\right\}
\Psi  \label{Breit}
\end{eqnarray}
in 1929 as a correction to an earlier defective equation which he called the
Eddington-Gaunt equation.

The Eddington-Gaunt equation \cite{edgnt} has the form 
\begin{eqnarray}
E\Psi =\left\{ \vec{\alpha}_1\cdot \vec{p}_1+\beta _1m_1+\vec{\alpha}_2\cdot 
\vec{p}_2+\beta _2m_2-\frac \alpha r(1-\vec{\alpha}_1\cdot \vec{\alpha}%
_2)\right\} \Psi .
\end{eqnarray}
This equation was defective because it failed to include the
semirelativistic electrodynamic interaction of Darwin \cite{dar}, together
with the Coulomb interaction in the combination

\begin{eqnarray}
-\frac \alpha r[1-\frac 12(\vec{v}_1\cdot \vec{v}_2+\vec{v}_1\cdot \hat{r}%
\vec{v}_2\cdot \hat{r})].
\end{eqnarray}
This structure arises from the semirelativistic expansion of the Green
function, including retardative terms through order $\frac 1{c^2}$.

\[
\int \int J_1GJ_2=\int d\tau _1\int d\tau _2\dot{x}_1^\mu \dot{x}_{2\mu
}\delta [(x_1-x_2)^2] 
\]
\begin{eqnarray}
\to -\frac \alpha r[1-\frac 12(\vec{v}_1\cdot \vec{v}_2+\vec{v}_1\cdot \hat{r%
}\;\vec{v}_2\cdot \hat{r})].
\end{eqnarray}
Replacing velocities in the Darwin interaction by $\vec{\alpha}$'s led Breit
to terms missing from Eddington's approach.

In modern language the Eddington-Gaunt interaction is most closely connected
to QED in the Feynman gauge while the Breit interaction is most closely
connected to QED in the Coulomb gauge. Eddington's incorrect implementation
of the Feynman gauge produces the wrong QED spectrum while Breit's correct
implementation of the Coulomb gauge produces the correct QED spectrum for
positronium and muonium.

Since the Breit equation is not covariant and not a well defined wave
equation it must be handled by semirelativistic perturbative methods. For
example, for QED it is standardly rearranged as the semirelativistic
elaboration of the Schrodinger Equation:

\begin{eqnarray}
H\psi =w\psi .
\end{eqnarray}
($w$ is the total c.m. energy) in which the Hamiltonian is the first-order
perturbative form 
\[
H=(m_1+{\frac{\vec{p}^2}{2m_1}}-{\frac{(\vec{p}^2)^2}{8m_1^3}})+(m_2+{\frac{%
\vec{p}^2}{2m_2}}-{\frac{(\vec{p}^2)^2}{8m_2^3}})-\frac \alpha r+ 
\]
\[
-\alpha \big(-[{\frac{p^2}{m_1m_2}}{\frac 1r}+{\frac 1{2m_2m_2r}}\vec{p}%
\cdot (1-\hat{r}\hat{r})\cdot \vec{p}]_{ordered} 
\]
\[
-{\frac 12}({\frac 1{m_1^2}}+{\frac 1{m_2^2}})\delta (\vec{r})-{\frac 14}{%
\frac{\vec{L}}{r^3}}\cdot [({\frac 1{m_1^2}}+{\frac 2{m_1m_2}})\vec{\sigma}%
_1+({\frac 1{m_2^2}}+{\frac 2{m_1m_2}})\vec{\sigma}_2] 
\]
\begin{eqnarray}
+{\frac 1{4m_1m_2}}(-{\frac{8\pi }3}\vec{\sigma}_1\cdot \vec{\sigma}_2\delta
(\vec{r})+{\frac{\vec{\sigma}_1\cdot \vec{\sigma}_2}{r^3}}-{\frac{3\vec{%
\sigma}_1\cdot \vec{r}\vec{\sigma}_2\cdot \vec{r}}{r^5}})\big).
\end{eqnarray}

\subsection{Manifestly Covariant Two-Body Dirac Equations}

In the 1970's, several authors used Dirac's constraint mechanics to attack
the relativistic two-body problem at its classical roots \cite{cnstr}
successfully evading the so-called no interaction theorem \cite{nogo}. Using
this method, the present authors extended the constraint approach to pairs
of spin one half particles to obtain two-body quantum bound state equations
that correct \cite{cra96} the defects in the Breit equation and more
importantly in the ladder approximation to the Bethe-Salpeter equation,
exorcising quantum ghosts by covariantly controlling the relative time
variable. Those equations are the two-body Dirac equations of constraint
dynamics \cite{cra82}. They possess a number of important features some of
which are unique and which correct defects in patchwork approaches. They are
manifestly covariant while yielding simple three-dimensional
Schr\"{o}dinger-like forms similar to those of their nonrelativistic
counterparts. Their spin dependence is not put in by hand, as in patchwork
approaches, but is determined naturally by the Dirac-like structure of the
equations. These equations have passed numerous tests showing that they
reproduce correct QED perturbative results when solved nonperturbatively.
They thus qualify as bona fide wave equations. In addition, the Dirac forms
of the equations automatically make unnecessary the ad hoc introduction of
cutoff parameters. The relativistic potentials appearing in these equations
are related directly to the interactions of perturbative quantum field
theory or (for QCD) may be introduced semiphenomenologically. In QCD one can
regard these manifestly covariant equations as an anticipation of those that
may eventually emerge from lattice gauge theory as applied to meson
spectroscopy.

As in the ordinary one-body Dirac equation Eq.(\ref{obd}), for particles interacting
through world vector and scalar interactions the two-body Dirac equations
equations take the general minimal coupling form 
\begin{mathletters}
\label{two}
\begin{eqnarray}
{\cal S}_1\psi \equiv \gamma _{51}(\gamma _1\cdot (p_1-\tilde{A}_1)+m_1+%
\tilde{S}_1)\psi =0
\end{eqnarray}
\begin{eqnarray}
{\cal S}_2\psi \equiv \gamma _{52}(\gamma _2\cdot (p_2-\tilde{A}_2)+m_2+%
\tilde{S}_2)\psi =0.
\end{eqnarray}
These equations provide a non-perturbative framework for extrapolating
perturbative field theoretic results into the highly relativistic regime of
bound light quarks, in a quantum mechanically well defined way. That
framework incorporates of three main properties:

\noindent I: Exact Lorentz covariance,

\noindent II: Minimal interaction structure

\noindent III: Compatibility of the two-equations (which leads to a
relativistic 3rd law, covariantly restricting the relative momentum and
energy while correctly structuring spin-dependent recoil): 
\end{mathletters}
\begin{eqnarray}
\lbrack {\cal S}_1,{\cal S}_2]\psi =0.
\end{eqnarray}
The satisfaction of this requirement originates in part from the presence of
supersymmetries in the (pseudo-)classical limit of the Two-Body Dirac
Equations. The potentials in these equations for four-vector and world
scalar interactions are intimately connected to those of Wheeler-Feynman
Electrodynamics (and its scalar counterpart) \cite{cra87,cra82,cra94} and
have been obtained systematically from perturbative quantum field theory 
\cite{saz85,cra88}.

For vector interactions alone, their momentum and spin dependences take the
simple ''hyperbolic'' forms \cite{cra86,cra97} 
\begin{mathletters}
\begin{eqnarray}
\tilde{A}_1=[1-{\rm cosh}({\cal G})]p_1+{\rm sinh}({\cal G})p_2-\frac i2%
(\partial {\rm e}^{{\cal G}}\cdot \gamma _2)\gamma _2
\end{eqnarray}
\begin{eqnarray}
\tilde{A}_2=[1-{\rm cosh}({\cal G})]p_2+{\rm sinh}({\cal G})p_1+\frac i2%
(\partial {\rm e}^{{\cal G}}\cdot \gamma _1)\gamma _1
\end{eqnarray}
in which 
\end{mathletters}
\begin{eqnarray}
\label{int}
{\cal G}=-{\frac 12}ln(1-2{\cal A}/w)
\end{eqnarray}
(with $w$ the total c.m. energy). We originally found the logarithm form in
a derivation from the Wheeler-Feynman classical field theory. In fact, in
quantum electrodynamics that form turns out to embody an eikonal summation
of ladder and cross-ladder diagrams.

In Eqs.(\ref{int}), the invariant ${\cal A}$ is a function of the covariant spacelike
particle separation 
\begin{eqnarray}
x_{\perp }^\mu =x^\mu +\hat{P}^\mu (\hat{P}\cdot x)
\end{eqnarray}
perpendicular to the total four-momentum, $P$. ($\hat{P}\equiv {\frac Pw}$
is a time-like unit vector.) Its appearance signifies that the dynamics is
independent of the relative time in the c.m system.

For lowest order electrodynamics, 
\begin{eqnarray}
{\cal A}={\cal A}(x_{\perp })=-{\frac \alpha r}
\end{eqnarray}
in which 
\begin{eqnarray}
r\equiv \sqrt{x_{\perp }^2}.
\end{eqnarray}
The form of the covariant spin-dependent terms and the fact that ${\cal A}$
depends on $x_{\bot }$ are consequences of compatibility of the two Dirac
equations ($[{\cal S}_1,{\cal S}_2]\psi =0$). In quark-model calculations,
the invariant ${\cal A}$ and its counterpart for the scalar interaction are
chosen on semiphenomenological grounds.

These two-body Dirac equations bypass most of the difficulties of the
Bethe-Salpeter equation that arise from the presence relative time and
energy variables, and yield a three-dimensional but manifestly covariant
rearrangement of the Bethe-Salpeter equation. The three dimensional
character is partially embodied in the invariant $r$ which reduces to the
interparticle separation only in the c.m. system. The fact that the
interaction is instantaneous in the c.m. system is a direct consequence of
the compatibility of the two equations. It is not an ad hoc restriction
imposed on the equation as is done for various instantaneous approximations
of the Bethe-Salpeter equation \cite{cra88}.

Just as Dirac arrived at his equation by ''taking the square root of the
Klein-Gordon equation'' so these equations can be derived by rigorously
''taking the square root'' of the corresponding compatible ''two body
Klein-Gordon equations'' \cite{cra82}

\bgm\bg
{\cal S}_1\equiv \gamma _{51}(\gamma _1\cdot (p_1-\tilde{A}_1)+m_1+\tilde{S}%
_1)=``\sqrt{(p_1-A_1)^2+(m_1+S_1)^2+...}" \en
\bg
{\cal S}_2\equiv \gamma _{51}(\gamma _1\cdot (p_1-\tilde{A}_1)+m_1+\tilde{S}%
_1)=``\sqrt{(p_1-A_1)^2+(m_1+S_1)^2+...}". 
\en\enm
It is in this sense that it is most natural to call 
Eqs.(\ref{two}) ''Two-Body Dirac
equations'' .

These equations are not only manifestly covariant but are also
quantum-mechanically well-defined. That is, their covariant
Schr\"{o}dinger-like forms for the effective particle of relative motion 
\begin{eqnarray}
\big(p^2+\Phi _w(\sigma _1,\sigma _2,{\cal A}(r),S(r))\big)\psi =b^2(w)\psi
\end{eqnarray}
in which 
\begin{eqnarray}
b^2(w)={\frac 1{4w^2}}(w^4-2(m_1^2+m_2^2)w^2+(m_1^2-m_2^2)^2),
\end{eqnarray}
can be solved nonperturbatively for both QED and QCD bound state
calculations since every term in $\Phi _w(\sigma _1,\sigma _2,{\cal A}%
(r),S(r))$ is well defined (less singular than $-1/4r^2$) \cite{cra84,cra88}%
. Furthermore, recent work has shown that the Schr\"{o}dinger-like forms can
be transformed into equations that, like their nonrelativistic counterparts,
involve at most 2 coupled wave functions \cite{saz94,cra98} even when
non-central tensor forces or spin-difference-orbit interactions are present.
Note that in the decoupled form (or any other convenient form) the specific
forms of the spin dependent potentials are dictated (through the reduction
process) by the interaction structure of the original Dirac equation Eq.\ref
{two} and are not put in by hand.

We have checked the nonperturbative validity of these equations as well
defined wave equations by solving them analytically and numerically to
obtain the standard fine and hyperfine spectra of QED. For example, we
obtained an exact spectral solution for the singlet positronium system $%
{\cal A}=-\alpha /r$ 
\begin{eqnarray}
w=m\sqrt{2+2/\sqrt{1+{\frac{\alpha ^2}{(n+\sqrt{(l+{\frac 12})^2-\alpha ^2}%
-l-{\frac 12})^2}}}}
\end{eqnarray}
of the fully coupled system of 16-component equations \cite{cra86,cra92} 
\[
{\cal S}_1\psi ={\cal S}_2\psi =0. 
\]
Such validation we claim ought to be required of all candidate equations for
nonperturbative quark model calculations. No others on that list of 
60,
including the Salpeter reduction (the no-pair Breit equation) and
the Blankenbecler-Sugar equation, have yet met
this demand. If not required to meet this demand two body formalisms may
lead to possibly spurious nonperturbative predictions. For example Spence
and Vary \cite{vry} have shown that in their treatment, the no-pair Breit 
equation and the Blankenbecler-Sugar equation predict low energy
electron-positron resonances between 1.4 and 2.2 MeV in Bhabba scattering.
Such states have not been observed in low energy electron-positron
collisions nor are they predicted by our equations. With no rigorous check
on nonperturbative solutions for ordinary QED bound states, how can the QCD
results of such treatments be trusted?

Given a static potential model $V=V(r)$ for the quark-antiquark interaction
we can incorporate it in a covariant way into our equations by \cite
{cra84,cra88}

\noindent a) replacing nonrelativistic $r$ by $\sqrt{x_\perp^2}$

\noindent b) parcelling out the static potential $V$ into the invariant
functions ${\cal A}(r)$ and $S(r)$. This step remains a partially
phenomenological one. However, in our approach once ${\cal A}$ and $S$ are
fixed so are all the accompanying spin dependences. One cannot adjust the
various parts independently, as is done in many approaches 
\cite{isgur}
which just add plausible potential energy terms to two-body relativistic
kinetic energy operators.

\subsection{Comparison of Two-Body Dirac with Breit, Eddington-Gaunt and
other Approaches}

Since the Two-Body Dirac Equations make quantum mechanical sense they can be
solved nonperturbatively in QCD bound state calculations, just like the
ordinary one-body Dirac equation (to which they reduce in the limit that
either particle becomes infinitely massive. The well-defined potential
structures function as a natural smoothing mechanism that yields the correct
spectrum while avoiding singular effective potentials like delta functions
that appear in the Breit reductions and most competing approaches. For
example, the Pauli, i.e. Schr\"{o}dinger-like forms of our equations make
quantum mechanical sense in the strong potential, nonperturbative regime
where relativistic effects of the wave operator on the wave function are not
negligible. This claim is easiest to see by examining the connection \cite
{cra84} of the main spin-spin term in our equation with that of Breit.

\begin{eqnarray}
({\rm Two-Body\ Dirac})\ \ -{\frac{1}{6}}\sigma_1\cdot\sigma_2\partial^2
ln(1-{\frac{2{\cal A}}{w}})\to {\frac{1}{3}}{\frac{\sigma_1\cdot\sigma_2
\partial^2{\cal A}}{m_1+m_2}}\ \ {\rm Breit\ }
\end{eqnarray}

For ${\cal A}$'s that have singular short range behaviors like $-\alpha /r$
(QED) and $8\pi /27rlnr$ (QCD) the weak potential form which appears in the
reduced Breit equation and most patchwork approaches can only be used in a
perturbative calculation. The two-body Dirac form on the left can be used
nonperturbatively with the logarithm term providing a natural smoothing
mechanism avoiding the necessity of introducing singularity softening
parameters in phenomenological approaches \cite{cra84,cra88}.

For QED, the Two-Body Dirac equations work naturally in the covariant
Feynman gauge, organizing the diagrammatic summation in a more efficient way
than does the BSE. They naturally produce an equation closely connected in
form to the defective equation of Eddington but yielding a correct spectrum.
They achieve this result by effectively summing an infinite number of ladder
and cross ladder diagrams in a kind of eikonal approximation \cite{sazj}.
Can we see this effect directly? Since they form a compatible pair they can
be combined in any number of equivalent ways, in particular in a Breit-like
form \cite{cra94,cra96,cra97,saz95}:

\begin{eqnarray}
w\Psi =\left\{ \vec{\alpha}_1\cdot \vec{p}_1+\beta _1m_1+\vec{\alpha}_2\cdot 
\vec{p}_2+\beta _2m_2+w(1-\exp \big[-{\cal G}(x_{\bot })(1-\vec{\alpha}%
_1\cdot \vec{\alpha}_2)\big])\right\} \Psi \text{ ,}
\end{eqnarray}
\begin{eqnarray}
{\cal G}=\ -{\frac 12}ln(1+{\frac{2\alpha }{rw}})=-{\frac \alpha {wr}}+...
\end{eqnarray}
Comparing this to Eddington and Gaunt's 
\begin{eqnarray}
w\Psi =\left\{ \vec{\alpha}_1\cdot \vec{p}_1+\beta _1m_1+\vec{\alpha}_2\cdot 
\vec{p}_2+\beta _2m_2-\frac \alpha r(1-\vec{\alpha}_1\cdot \vec{\alpha}%
_2)\Psi \right\} ,
\end{eqnarray}
we see that by effectively stopping at lowest order, the Eddington-Gaunt
equation was doomed to failure.

In detail, to all orders in the potential, our equations produce \cite{cra97}
\begin{eqnarray}
w(1-\exp [-{\cal G}(x_{\bot })(1-\vec{\alpha}_1\cdot \vec{\alpha}_2)])
\end{eqnarray}
\begin{eqnarray}
={\cal A}(1-\vec{\alpha}_1\cdot \vec{\alpha}_2)-\frac{{\cal A}^2}w(1-\vec{%
\sigma}_1\cdot \vec{\sigma}_2)-{\frac{{\cal A}^3}{w^2}}{\frac 1{(1-2{\cal A}%
/w)}}(1-\gamma _{51}\gamma _{52}+\vec{\alpha}_1\cdot \vec{\alpha}_2-\vec{%
\sigma}_1\cdot \vec{\sigma}_2)\text{ .}
\end{eqnarray}
>From the point of view of the Breit form of our equations this means that
not only do they contain vector interactions but other covariant
interactions as well (e.g. pseudovector, scalar and pseudoscalar
interactions), which plausibly can be viewed as originating in the sorts of
products of vector interactions that occur in multiparticle exchange. Using
a transformation due to Schwinger, we have shown elsewhere that the extra
terms that remain in the weak potential limit the are canonically equivalent
to Breit's retardative terms \cite{cra97}.

\subsection{Connection to Quantum Field Theory}

The invariant forms ${\cal A}$ and $S$ in the Two-Body Dirac Equations 
Eq.(\ref{two})
may be systematically obtained from the corresponding quantum field theories 
\cite{saz85,cra88}. The connection is 
\begin{eqnarray}
\Phi _w(\sigma _1,\sigma _2,{\cal A}(r),S(r))=\pi i\delta (\hat{P}\cdot p)%
{\cal K}(1+\bar{{\cal K}})^{-1}
\end{eqnarray}
giving the quasipotential in terms of the Bethe-Salpeter kernel ${\cal K}$
and its projection 
\begin{eqnarray}
\bar{{\cal K}}={\bf G}{\cal K}
\end{eqnarray}
in which 
\begin{eqnarray}
{\bf G}\equiv \big ({\frac 1{p_1^2+m_1^2-i0}}{\frac 1{p_2^2+m_2^2-i0}}-\pi
i\delta (P\cdot p){\frac w{p_{\perp }^2-b^2(w)-i0}}\big )
\end{eqnarray}
is the difference between forms of two-body propagators as given by the
Bethe-Salpeter equation on the one hand and the constraint equations on the
other (with the relativistic third law delta function.) ($p$ is the relative
momentum and $P$ is the total momentum.) This connection is a sort of
covariant version of the three-dimensional Lippman-Schwinger equation. The
difference is that in this equation, derived by Sazdjian \cite{saz85} as
''quantum mechanical transform of the Bethe-Salpeter equation'', the
potential follows from the irreducible scattering matrix rather than the
other way around as done in the nonrelativistic case. The present authors
have also derived the effective potentials from classical field theory in
the form of Fokker-Tetrode actions through comparison with the classical
limit of the constraint equations\cite{fktr,cra94}.

\section{Two-Body Dirac Equations and Meson Spectroscopy}

We have formulated a constraint version of the naive quark model for mesons
by using the static Adler-Piran quark-antiquark potential \cite{adlr},
(covariantly reinterpreted) in our Two-Body Dirac Equations. Adler and Piran
obtained their static quark potential from an effective non-linear field
theory derived from QCD. It has the general form 
\begin{eqnarray}
V_{AP}(r)=\Lambda (U(\Lambda r)+U_0)\ (={\cal A}+S).
\end{eqnarray}
Since their potential is nonrelativistic it cannot distinguish between world
scalar and vector potentials, simply representing the effect of their sum in
the nonrelativistic limit. It incorporates asymptotic freedom analytically
through 
\begin{eqnarray}
\Lambda U(\Lambda r<<1)\sim 1/(rln\Lambda r)
\end{eqnarray}
and linear confinement through 
\begin{eqnarray}
\Lambda U(\Lambda r>>1)\sim \Lambda ^2r.
\end{eqnarray}
At long distances their potential includes not only the linear confinement
piece but also subdominant logarithm terms among others 
\begin{eqnarray}
V_{AP}(r)=\Lambda (c_1x+c_2ln(x)+c_3/\sqrt{x}+c_4/x+c_5),\ \ x\equiv \Lambda
r>2.
\end{eqnarray}
The $c_is$ are given by the Adler-Piran leading log-log model. The realistic
Adler-Piran potential or ones like it such as the Richardson potential \cite
{rich}, fail miserably for light mesons when used in the nonrelativistic
Schr\"{o}dinger equation \cite{cra81}. Exact covariance is essential to
handle the light mesons if one insists, as we do, on using potentials
closely tied to QCD.

We apportion the potential between the relativistic invariants $S$ and $%
{\cal A}$ that determine the scalar and vector potentials according to the
scheme 
\begin{mathletters}
\begin{eqnarray}
{\cal A}=exp(-\beta r)[V_{AP}-{\frac{c_4}{r}}]+{\frac{c_4}{r}}+{\frac{e_1e_2%
}{r}} ,  \label{vctr}
\end{eqnarray}
\begin{eqnarray}
S=V_{AP}+{\frac{e_1e_2}{r}}-{\cal A}.  \label{sclr}
\end{eqnarray}
In this way we impose the requirement that at short distance the potential
is strictly vector while at long distance the confining portion is scalar
with Coulomb vector portions. The relativistic invariance of $S$ and ${\cal A%
}$ follows by reinterpreting the variable $r$ as $r\equiv \sqrt{x_\perp^2}.$

Our quark model is a naive quark model in that we ignore flavor mixing and
the effects on the bound state energies of decays. The results we obtain,
using the same potentials for all of the mesons, are spectrally quite
accurate, from the heaviest upsilonium states to the lowly pion.

{\bf TABLE I - MESON MASSES FROM COVARIANT CONSTRAINT TWO-BODY DIRAC
EQUATIONS} \halign{#\hfil&\qquad\hfil#&\qquad\hfil#\cr
NAME & EXP. & THEORY\cr\cr
$\Upsilon : b \overline b \ 1^3S_1$ & 9.460( 0.2)& 9.453( 0.6)\cr
$\Upsilon : b \overline b \ 1^3P_0$ & 9.860( 1.3)& 9.842( 1.4)\cr
$\Upsilon : b \overline b \ 1^3P_1$ & 9.892( 0.7)& 9.889( 0.1)\cr
$\Upsilon : b \overline b \ 1^3P_2$ & 9.913( 0.6)& 9.921( 0.5)\cr
$\Upsilon : b \overline b \ 2^3S_1$ & 10.023( 0.3)& 10.022( 0.0)\cr
$\Upsilon : b \overline b \ 2^3P_0$ & 10.232( 0.6)& 10.227( 0.2)\cr
$\Upsilon : b \overline b \ 2^3P_1$ & 10.255( 0.5)& 10.257( 0.0)\cr
$\Upsilon : b \overline b \ 2^3P_2$ & 10.269( 0.4)& 10.277( 0.8)\cr
$\Upsilon : b \overline b \ 3^3S_1$ & 10.355( 0.5)& 10.359( 0.1)\cr
$\Upsilon : b \overline b \ 4^3S_1$ & 10.580( 3.5)& 10.614( 0.9)\cr
$\Upsilon : b \overline b \ 5^3S_1$ & 10.865( 8.0)& 10.826( 0.2)\cr
$\Upsilon : b \overline b \ 6^3S_1$ & 11.019( 8.0)& 11.013( 0.0)\cr
$B: b \overline u \ 1^1S_0$ & 5.279( 1.8)& 5.273( 0.1)\cr
$B: b \overline d \ 1^1S_0$ & 5.279( 1.8)& 5.274( 0.1)\cr
$B^*: b \overline u \ 1^3S_1$ & 5.325( 1.8)& 5.321( 0.1)\cr
$B_s: b \overline s \ 1^1S_0$ & 5.369( 2.0)& 5.368( 0.0)\cr
$B_s: b \overline s \ 1^3S_1$ & 5.416( 3.3)& 5.427( 0.1)\cr
$\eta_c : c \overline c \ 1^1S_0$ & 2.980( 2.1)& 2.978( 0.0)\cr
$\psi: c \overline c \ 1^3S_1$ & 3.097( 0.0)& 3.129( 12.6)\cr
$\chi_0: c \overline c \ 1^1P_1$ & 3.526( 0.2)& 3.520( 0.4)\cr
$\chi_0: c \overline c \ 1^3P_0$ & 3.415( 1.0)& 3.407( 0.4)\cr
$\chi_1: c \overline c \ 1^3P_1$ & 3.510( 0.1)& 3.507( 0.2)\cr
$\chi_2: c \overline c \ 1^3P_2$ & 3.556( 0.1)& 3.549( 0.6)\cr
$\eta_c : c \overline c \ 2^1S_0$ & 3.594( 5.0)& 3.610( 0.1)\cr
$\psi: c \overline c \ 2^3S_1$ & 3.686( 0.1)& 3.688( 0.1)\cr
$\psi: c \overline c \ 1^3D_1$ & 3.770( 2.5)& 3.808( 2.0)\cr
$\psi: c \overline c \ 3^3S_1$ & 4.040( 10.0)& 4.081( 0.2)\cr
$\psi: c \overline c \ 2^3D_1$ & 4.159( 20.0)& 4.157( 0.0)\cr
$\psi: c \overline c \ 3^3D_1$ & 4.415( 6.0)& 4.454( 0.4)\cr
$D: c \overline u \ 1^1S_0$ & 1.865( 0.5)& 1.866( 0.0)\cr
$D: c \overline d \ 1^1S_0$ & 1.869( 0.5)& 1.873( 0.1)\cr
$D^*: c \overline u \ 1^3S_1$ & 2.007( 0.5)& 2.000( 0.4)\cr
$D^*: c \overline d \ 1^3S_1$ & 2.010( 0.5)& 2.005( 0.3)\cr
$D^*: c \overline u \ 1^3P_1$ & 2.422( 1.8)& 2.407( 0.6)\cr
$D^*: c \overline d \ 1^3P_1$ & 2.428( 1.8)& 2.411( 0.5)\cr
$D^*: c \overline u \ 1^3P_2$ & 2.459( 2.0)& 2.382( 11.3)\cr
$D^*: c \overline d \ 1^3P_2$ & 2.459( 4.0)& 2.386( 3.5)\cr
$D_s: c \overline s \ 1^1S_0$ & 1.968( 0.6)& 1.976( 0.5)\cr
$D_s^*: c \overline s \ 1^3S_1$ & 2.112( 0.7)& 2.123( 0.9)\cr
$D_s^*: c \overline s \ 1^3P_1$ & 2.535( 0.3)& 2.511( 6.2)\cr
$D_s^*: c \overline s \ 1^3P_2$ & 2.574( 1.7)& 2.514( 9.6)\cr
$K: s \overline u \ 1^1S_0$ & 0.494( 0.0)& 0.492( 0.0)\cr
$K: s \overline d \ 1^1S_0$ & 0.498( 0.0)& 0.492( 0.4)\cr
$K^*: s \overline u \ 1^3S_1$ & 0.892( 0.2)& 0.910( 0.6)\cr
$K^*: s \overline d \ 1^3S_1$ & 0.896( 0.3)& 0.910( 0.3)\cr
$K_1: s \overline u \ 1^1P_1$ & 1.273( 7.0)& 1.408( 3.2)\cr
$K_0^*: s \overline u \ 1^3P_0$ & 1.429( 4.0)& 1.314( 0.7)\cr
$K_1: s \overline u \ 1^3P_1$ & 1.402( 7.0)& 1.506( 1.0)\cr
$K_2^*: s \overline u \ 1^3P_2$ & 1.425( 1.3)& 1.394( 0.5)\cr
$K_2^*: s \overline d \ 1^3P_2$ & 1.432( 1.3)& 1.394( 0.6)\cr
$K^*: s \overline u \ 2^1S_0$ & 1.460( 30.0)& 1.591( 0.2)\cr
$K^*: s \overline u \ 2^3S_1$ & 1.412( 12.0)& 1.800( 6.7)\cr
$K_2: s \overline u \ 1^1D_2$ & 1.773( 8.0)& 1.877( 0.8)\cr
$K^*: s \overline u \ 1^3D_1$ & 1.714( 20.0)& 1.985( 1.4)\cr
$K_2: s \overline u \ 1^3D_2$ & 1.816( 10.0)& 1.945( 1.3)\cr
$K_3: s \overline u \ 1^3D_3$ & 1.770( 10.0)& 1.768( 0.0)\cr
$K^*: s \overline u \ 3^1S_0$ & 1.830( 30.0)& 2.183( 1.4)\cr
$K_2^*: s \overline u \ 2^3P_2$ & 1.975( 22.0)& 2.098( 0.2)\cr
$K_4^*: s \overline u \ 1^3F_4$ & 2.045( 9.0)& 2.078( 0.1)\cr
$K_2: s \overline u \ 2^3D_2$ & 2.247( 17.0)& 2.373( 0.5)\cr
$K_5^*: s \overline u \ 1^3G_5$ & 2.382( 33.0)& 2.344( 0.0)\cr
$K_3^*: s \overline u \ 2^3F_3$ & 2.324( 24.0)& 2.636( 1.9)\cr
$K_4^*: s \overline u \ 2^3F_4$ & 2.490( 20.0)& 2.757( 1.6)\cr
$\phi: s \overline s \ 1^3S_1$ & 1.019( 0.0)& 1.033( 2.2)\cr
$f_0: s \overline s \ 1^3P_0$ & 1.370( 40.0)& 1.319( 0.0)\cr
$f_1: s \overline s \ 1^3P_1$ & 1.512( 4.0)& 1.533( 0.3)\cr
$f_2: s \overline s \ 1^3P_2$ & 1.525( 5.0)& 1.493( 0.3)\cr
$\phi: s \overline s \ 2^3S_1$ & 1.680( 20.0)& 1.850( 0.8)\cr
$\phi: s \overline s \ 1^3D_3$ & 1.854( 7.0)& 1.848( 0.0)\cr
$f_2: s \overline s \ 2^3P_2$ & 2.011( 69.0)& 2.160( 0.1)\cr
$f_2: s \overline s \ 3^3P_2$ & 2.297( 28.0)& 2.629( 1.6)\cr
$\pi: u\overline d \ 1^1S_0$ & 0.140( 0.0)& 0.144( 0.2)\cr
$\rho: u\overline d \ 1^3S_1$ & 0.767( 1.2)& 0.792( 0.1)\cr
$b_1: u\overline d \ 1^1P_1$ & 1.231( 10.0)& 1.392( 2.1)\cr
$a_0: u\overline d \ 1^3P_0$ & 1.450( 40.0)& 1.491( 0.0)\cr
$a_1: u\overline d \ 1^3P_1$ & 1.230( 40.0)& 1.568( 0.7)\cr
$a_2: u\overline d \ 1^3P_2$ & 1.318( 0.7)& 1.310( 0.0)\cr
$\pi: u\overline d \ 2^1S_0$ & 1.300( 100.0)& 1.536( 0.1)\cr
$\rho: u\overline d \ 2^3S_1$ & 1.465( 25.0)& 1.775( 1.4)\cr
$\pi_2: u\overline d \ 1^1D_2$ & 1.670( 20.0)& 1.870( 0.9)\cr
$\rho: u\overline d \ 1^3D_1$ & 1.700( 20.0)& 1.986( 1.9)\cr
$\rho_3: u\overline d \ 1^3D_3$ & 1.691( 5.0)& 1.710( 0.0)\cr
$\pi: u\overline d \ 3^1S_0$ & 1.795( 10.0)& 2.166( 7.9)\cr
$\rho: u\overline d \ 3^3S_1$ & 2.149( 17.0)& 2.333( 0.7)\cr
$\rho_4: u\overline d \ 1^3F_4$ & 2.037( 26.0)& 2.033( 0.0)\cr
$\pi_2: u\overline d \ 2^1D_2$ & 2.090( 29.0)& 2.367( 0.5)\cr
$\rho_3: u\overline d \ 2^3D_3$ & 2.250( 45.0)& 2.305( 0.0)\cr
$\rho_5: u\overline d \ 1^3G_5$ & 2.330( 35.0)& 2.307( 0.0)\cr
$\rho_6: u\overline d \ 1^3H_6$ & 2.450( 130.0)& 2.547( 0.0)\cr
$\chi^2$ & ---& 101.0\cr} (The numbers in parentheses represent experimental
uncertainties and $\chi^2$ contributions for each meson.)

The heavy upsilon fits to the ground and excited states are due more to the
specific form of the Adler-Piran potential than to the relativistic features
of the equations, since the heavy quark motions are nearly nonrelativistic.
The spin-orbit splittings arise from our apportionment of the Adler
potential into scalar and vector parts. The good fits to the heavy $B$
mesons are due primarily to the fact that in the infinite quark mass limit
our equations reduce to the one body Dirac equation.

For charmonium, the semirelativistic effects of the formalism become
important, the higher order relativistic effects becoming more important in
the $D$ and then $\phi $ mesons. The only significant weakness in our fits
(due to limitations of apportioning our potential only between scalar and
vector potentials) manifests itself at the $LS$ multiplet level for the
lighter mesons. For those states, partial inversion takes place.

However, the equations work well in the extreme relativistic domain of the
very light mesons and their lower excitations - ($K,K^{*},\pi ,\rho $) and
first radial excitations.

In our equations, the pion is a Goldstone boson in the sense that its mass
tends toward zero numerically in the limit in which the quark mass
numerically goes toward zero. This may be seen in the accompanying plot.
Note that the $\rho $ meson mass approaches a finite value in the chiral
limit. This also holds true for the excited pion states.

\vspace*{8.5cm}
\epsfxsize=300pt
\includegraphics{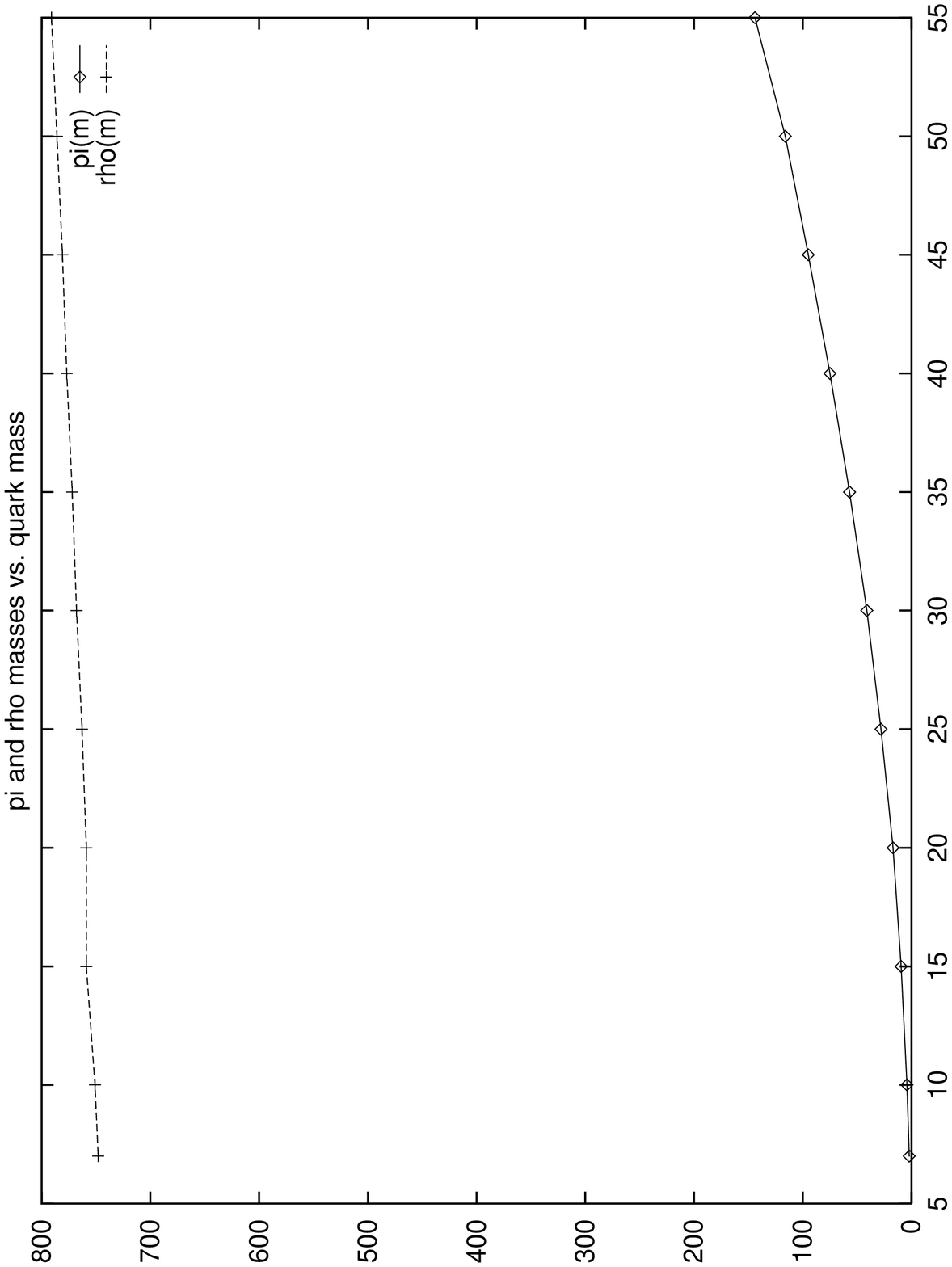}
\vspace*{-.2cm}\hspace*{2.5cm}
\begin{minipage}[t]{14cm}{{\bf Fig.\ 1.}$\pi$ and $\r$ masses versus quark mass (in MeV)}
\end{minipage}

Our results have room for improvement in the spin-orbit splittings of
the light mesons and other orbital and radial excitations. This is a
consequence of our assumption that at long distance the confining
interaction is pure scalar.

With just two parametric functions, (amounting to an ad hoc division of the
Adler potential into vector and scalar parts), we are able to obtain a fit
about as good as that obtained by Godfrey and Isgur \cite{isgur}, who
use six parametric functions, basically one for each type of spin
dependence. The Godfrey and Isgur results seem more accurate than ours for
the spin-orbit splitting of the light mesons, likely due to their use of
different parameters for each spin-dependent part. Note that they must also
introduce parameters that smooth the potential, a procedure that is not
necessary in the two-body Dirac equations because of the connection of their
structure to that of the one-body Dirac equation.

\section{ Two-Body Dirac Equations for General Covariant Interactions}

One may generalize the interactions in the Two-Body Dirac Equations to ones
other than vector and scalar. How does one determine the forms of these
equations for arbitrary interactions? We use this opportunity to discuss the
role played by supersymmetry in the origins of the constraint formalism for
two-body Dirac equations \cite{cra82,cra87}.

\subsection{The Role of Supersymmetry}

We first rewrite one-body Dirac equation in terms of theta matrices defined
by 
\bgm
\begin{eqnarray}
\theta ^\mu \equiv i\sqrt{\frac 12}\gamma _5\gamma _{,}^\mu \ \mu =0,1,2,3,\
\ \theta _5\equiv i\sqrt{\frac 12}\gamma _5,
\end{eqnarray}
\enm
\begin{eqnarray}
\lbrack \theta ^\mu ,\theta ^\nu ]_{+}=-\eta ^{\mu \nu },\ \ [\theta
_5,\theta ^\mu ]_{+}=0,\ \ [\theta _5,\theta _5]_{+}=-1.
\end{eqnarray}\enm
In terms of these the free Dirac equation and its Klein-Gordon square become 
\begin{eqnarray}
{\cal S}\psi \equiv (p\cdot \theta +m\theta _5)\psi =0
\end{eqnarray}
in which ${\cal S}$ is literally the ``operator square root'' of Einstein's
mass shell condition: 
\begin{eqnarray}
{\cal S}^2\psi =-{\frac 12}(p^2+m^2)\psi =0.
\end{eqnarray}
The supersymmetry is most easily revealed through examination of
''pseudoclassical'' mechanics - the ''correspondence-principle'' limit of
the Dirac equation. In that limit, these theta matrices become Grassmann
variables obeying ``pseudoclassical'' Berezin-Marinov brackets which upon
quantization generate the Dirac algebra \cite{brmn}: $\{\theta ^\mu ,\theta
^\nu \}=i\eta ^{\mu \nu },\{\theta _5,\theta ^\mu \}=0,\{\theta _5,\theta
_5\}=i.$

In this limit, the Dirac equation becomes a constraint imposed on both
bosonic ($p$) and fermionic ($\theta ,\theta _5$) variables: 
\begin{eqnarray}
{\cal S}\equiv (p\cdot \theta +m\theta _5)\approx 0
\end{eqnarray}
while the mass shell condition comes from the bracket 
\begin{eqnarray}
{\cal H}\equiv {\frac 1i}\{{\cal S},{\cal S}\}=p^2+m^2\approx 0.
\end{eqnarray}
We first examine the supersymmetry of the free Dirac constraint under
transformations generated by the self-abelian ${\cal G}$ defined by 
\begin{mathletters}

\begin{eqnarray}
{\cal G}=p\cdot \theta +\sqrt{-p^2}\theta _5,
\end{eqnarray}
\begin{eqnarray}
\{{\cal G},{\cal G}\}=0.
\end{eqnarray}
That is, 
\end{mathletters}
\begin{eqnarray}
\{{\cal G},{\cal S}\}\approx 0.
\end{eqnarray}
${\cal G}$ is a supersymmetry generator in the sense that it mixes bosonic
variables like the momentum with the Grassmann or fermionic variables. To
maintain this supersymmetry in the presence of interaction, we must
introduce interactions that have zero brackets with ${\cal G}$. The position
four-vector $x$ is not supersymmetric and in fact displays pseudoclassical
zitterbewegegung. However, the ``zitterbewegungless'' position variable 
\begin{eqnarray}
\tilde{x}^\mu =x^\mu +{\frac{i\theta ^\mu \theta _5}m}
\end{eqnarray}
is supersymmetric. But, because of the presence of $m$ in this variable,
this object must itself be modified in the presence of scalar interaction $%
M=m+S$. We accomplish this by generalizing this position variable to the
self-referent form 
\begin{eqnarray}
\tilde{x}^\mu =x^\mu +{\frac{i\theta ^\mu \theta _5}{M(\tilde{x})}}.
\end{eqnarray}
This is a supersymmetric position variable appropriate for scalar
interactions in the sense that 
\begin{eqnarray}
\{{\cal G},\tilde{x}\}\approx 0.
\end{eqnarray}
The supersymmetric constraints (both fermionic and bosonic) then become 
\begin{eqnarray}
{\cal S}=p\cdot \theta +M(\tilde{x})\theta _5\approx 0,\ {\frac 1i}\{{\cal S}%
,{\cal S}\}\equiv {\cal H}=p^2+M^2(\tilde{x})\approx 0.
\end{eqnarray}
Since $\theta _5^2=0$ ,the expansion of the self-referent form truncates 
\begin{eqnarray}
M(\tilde{x})=M(x)+{\frac{i\partial M(x)\cdot \theta \theta _5}{M(x)}}
\end{eqnarray}
so that the ${\cal S}$ constraint assumes the standard form 
\begin{eqnarray}
{\cal S}=p\cdot \theta +M(x)\theta _5\approx 0
\end{eqnarray}
while the mass-shell constraint becomes the familiar square 
\begin{eqnarray}
{\frac 1i}\{{\cal S},{\cal S}\}\equiv {\cal H}=p^2+M^2(\tilde{x}%
)=p^2+M^2(x)+2i\partial M(x)\cdot \theta \theta _5.
\end{eqnarray}
One arrives back at the Dirac equation and its square for scalar interaction
by replacing Grassmann variables with theta matrices and dynamical variables 
$x$ and $p$ with their operator forms. 
\begin{mathletters}

\begin{eqnarray}
{\cal S}\psi =[p\cdot \theta +M(x)\theta _5]\psi =0
\end{eqnarray}
\begin{eqnarray}
{\cal H}\psi =[p^2+M^2(x)+2i\partial M(x)\cdot \theta \theta _5]\psi =0.
\end{eqnarray}
These contain the usual spin-dependent corrections expected for scalar
interactions. We have gone through this exercise in order to show that this
structure is present in the usual result. The supersymmetry generated by $%
{\cal G}$ and realized through the presence of $\tilde{x}$ is a natural
feature of both the free Dirac equation and its standard form for external
scalar interaction. 

How does one implement this supersymmetry in the case of
two interacting particles? For two pseudoclassical free particles we begin
with 
\end{mathletters}
\begin{eqnarray}
{\cal S}_{i0}=p_i\cdot \theta _i+m_i\theta _{5i}\approx 0,\ i=1,2
\end{eqnarray}
in which the two sets of $\theta $'s are independent Grassmann variables so
that the mutual ${\cal S}_{i0}$ bracket vanishes strongly: 
\begin{eqnarray}
\{{\cal S}_{10},{\cal S}_{20}\}=0.
\end{eqnarray}
In the presence of interaction, we require the preservation of supersymmetry
for each spinning particle. For scalar interactions we accomplish this
through 
\begin{eqnarray}
m_i\to M_i(x_1-x_2)\to M_i(\tilde{x}_1-\tilde{x}_2)\equiv \tilde{M}_i,\ i=1,2
\end{eqnarray}
depending on a supersymmetric variable for each particle position 
\begin{eqnarray}
\tilde{x}_i^\mu =x_i^\mu +{\frac{i\theta _i^\mu \theta _{5i}}{\tilde{M}_i}}%
,\ i=1,2.
\end{eqnarray}
Note that the Grassmann Taylor expansions of the $\tilde{M}_i$ truncate.
Carrying out those expansions, we obtain the following two-body Dirac
constraints: 
\begin{mathletters}

\begin{eqnarray}
{\cal S}_1=p_1\cdot \theta _1+\tilde{M}_1\theta _{51}=p_1\cdot \theta
_1+M_1\theta _{51}-i{\frac{\partial M_1\cdot \theta _1\theta _{52}\theta
_{51}}{M_2}}\approx 0,
\end{eqnarray}
\begin{eqnarray}
{\cal S}_2=p_2\cdot \theta _2+\tilde{M}_2\theta _{52}=p_2\cdot \theta
_2+M_2\theta _{52}+i{\frac{\partial M_2\cdot \theta _2\theta _{51}\theta
_{52}}{M_1}}\approx 0.
\end{eqnarray}
whose ``squares'' become 
\end{mathletters}
\begin{eqnarray}
{\frac 1i}\{{\cal S}_i,{\cal S}_i\}\equiv {\cal H}_i=p_i^2+\tilde{M}_i^2,\
i=1,2.
\end{eqnarray}
One can show that all four ${\cal S}$ and ${\cal H}$'s are mutually
compatible provided that two other conditions are met. Straightforward
computation leads to 
\begin{eqnarray}
\{{\cal S}_1,{\cal S}_2\}=-p_1\cdot {\frac{\partial M_2}{M_1}}\theta
_{51}\theta _{52}-p_2\cdot {\frac{\partial M_1}{M_2}}\theta _{51}\theta
_{52}.
\end{eqnarray}
This bracket vanishes strongly provided that 
\begin{eqnarray}
\partial (M_1^2-M_2^2)=0,  \label{thrd}
\end{eqnarray}
(the relativistic ''third law'' condition), and 
\begin{eqnarray}
M_i=M_i(x_{\perp })
\end{eqnarray}
in which 
\begin{eqnarray}
x_{\perp }^\mu =(\eta ^{\mu \nu }-{\frac{(p_1+p_2)^\mu (p_1+p_2)^\nu }{%
(p_1+p_2)^2}})(x_1-x_2)_\nu 
\end{eqnarray}
which, eliminates covariantly the troublesome ''relative time''. Now we use
these results to rewrite the constraints in a form which we can easily
generalize to interactions other than scalar. The third law condition Eq.(%
\ref{thrd}) has the solution 
\begin{eqnarray}
M_1^2-M_2^2=m_1^2-m_2^2
\end{eqnarray}
which has the convenient invariant hyperbolic parametrization 
\begin{eqnarray}
M_1=m_1\ chL\ +m_2shL,\ M_2=m_2\ chL\ +m_1\ shL,
\end{eqnarray}
in which 
\begin{eqnarray}
L=L(x_{\perp }).
\end{eqnarray}
In this case just one invariant function suffices to define the two-body
interaction. For such interactions 
\begin{eqnarray}
{\cal H}_1-{\cal H}_2=p_1^2+\tilde{M}_1^2-p_2^2-\tilde{M}%
_2^2=p_1^2+m_1^2-p_2^2-m_2^2\approx 0
\end{eqnarray}
just as in the noninteracting case (and indeed also in the spinless
interacting case \cite{nospn,cra82}). When the potentials satisfy these
conditions, the Dirac-like constraints ${\cal S}_1$ and ${\cal S}_2$ are
compatible and generate a consistent ``pseudoclassical'' dynamics. Note that
the combination of the constraints corresponding to the difference of the
squares eliminates the relative energy in the c.m. rest frame. To see this
write the constraints in terms of the total four-momentum 
\begin{eqnarray}
P^\mu =p_1^\mu +p_2^\mu ;\ -P^2\equiv w^2;\hat{P}^\mu \equiv P^\mu /w,
\end{eqnarray}
and relative four-momentum 
\begin{eqnarray}
\ p^\mu =(\epsilon _2p_2^\mu -\epsilon _1p_2^\mu )/w;\ \epsilon _1+\epsilon
_2=w,\ \epsilon _1-\epsilon _2=(m_1^2-m_2^2)/w.
\end{eqnarray}
(Note that $p$ is canonically conjugate to $x_{\perp }$. The $\epsilon _i$'s
are the invariant c.m. energies of each of the (interacting) particles.)
Thus the constraint difference 
\begin{eqnarray}
{\cal H}_1-{\cal H}_2=-2P\cdot p\approx 0
\end{eqnarray}
places an invariant restriction on the relative energy. The other linearly
independent combination 
\begin{eqnarray}
(\epsilon _2{\cal H}_1+\epsilon _1{\cal H}_2)/w=p_{\perp }^2+\Phi
_w-b^2(w)\approx 0,
\end{eqnarray}
in which 
\begin{eqnarray}
b^2(w)=(w^4-2w^2(m_1^2+m_2^2)+(m_1^2-m_2^2)^2)/4w^2
\end{eqnarray}
incorporates exact relativistic two-body kinematics and governs the dynamics
through the quasipotential 
\begin{eqnarray}
\Phi _w\equiv \tilde{M}_1^2-m_1^2=\tilde{M}_2^2-m_2^2.
\end{eqnarray}
Just as in the one-body case, we canonically quantize this system by
replacing the Grassmann variables $\theta _{\mu i},\theta _{5i}\ i=1,2$ by
two mutually commuting sets of theta matrices, and 
\begin{eqnarray}
\{x^\mu ,p^\nu \}\to [x^\mu ,p^\nu ]=i\eta ^{\mu \nu }.
\end{eqnarray}
This leads to strongly compatible ($[{\cal S}_1,{\cal S}_2]=0$) two-body
Dirac equations (for scalar interactions) in minimal interaction form 
\begin{mathletters}

\begin{eqnarray}
{\cal S}_1\psi =(\theta _1\cdot p+\epsilon _1\theta _1\cdot \hat{P}%
+M_1\theta _{51}-i\partial L\cdot \theta _2\theta _{52}\theta _{51})\psi =0,
\end{eqnarray}
\begin{eqnarray}
{\cal S}_2\psi =(-\theta _2\cdot p+\epsilon _2\theta _2\cdot \hat{P}%
+M_2\theta _{52}+i\partial L\cdot \theta _1\theta _{52}\theta _{51})\psi =0,
\end{eqnarray}
in which 
\end{mathletters}
\begin{eqnarray}
\partial L={\frac{\partial M_1}{M_2}}={\frac{\partial M_2}{M_1}},
\end{eqnarray}
\begin{eqnarray}
M_1=m_1\ chL\ +m_2shL,\ M_2=m_2\ chL\ +m_1\ shL.
\end{eqnarray}
We see that in these coupled Dirac equations the remnants of pseudoclassical
supersymmetries are the extra spin dependent recoil corrections to the
ordinary one-body Dirac equations. Without those terms the two equations
would not be compatible. When $M_i=m_i+S_i,$ these extra terms vanish when
one of the particles becomes infinitely heavy.

\subsection{Hyperbolic form of the two-body Dirac equations for General
Covariant Interactions}

How do we introduce general interactions? We accomplish this by recasting
the minimal interaction forms of the two-body Dirac equations into a more
general form, one that generalizes the hyperbolic forms we encountered
above. Simple identities such as 
\begin{eqnarray}
ch^2(\Delta )-sh^2(\Delta )=1
\end{eqnarray}
lead to 
\begin{mathletters}

\begin{eqnarray}
{\cal S}_1\psi =(ch(\Delta ){\bf S}_1+sh(\Delta ){\bf S}_2)\psi =0,
\end{eqnarray}
\begin{eqnarray}
{\cal S}_2\psi =(ch(\Delta ){\bf S}_2+sh(\Delta ){\bf S}_1)\psi =0,
\end{eqnarray}
in which appear auxiliary constraints defined by 
\end{mathletters}
\begin{mathletters}

\begin{eqnarray}
{\bf S}_1\psi \equiv ({\cal S}_{10}ch(\Delta )+{\cal S}_{20}sh(\Delta ))\psi
=0,
\end{eqnarray}
\begin{eqnarray}
{\bf S}_2\psi \equiv ({\cal S}_{20}ch(\Delta )+{\cal S}_{10}sh(\Delta ))\psi
=0,
\end{eqnarray}
with 
\end{mathletters}
\begin{eqnarray}
\Delta =-\theta _{51}\theta _{52}L(x_{\perp }).
\end{eqnarray}
Note that in this form, the interaction enters only through an invariant
matrix function $\Delta $ with all other spin-dependence contained in the
kinetic free Dirac operators ${\cal S}_{10},{\cal S}_{20}$. One can show 
\cite{hyp,cra96} that both ${\cal S}_i$ and ${\bf S}_i$ constraints are
compatible for general $\Delta $: 
\begin{eqnarray}
\lbrack {\bf S}_1,{\bf S}_2]\psi =0\ {\rm and}\ [{\cal S}_1,{\cal S}_2]\psi
=0
\end{eqnarray}
provided only that 
\begin{eqnarray}
\Delta =\Delta (x_{\perp }).
\end{eqnarray}

\noindent For the polar interactions we find 
\begin{mathletters}

\begin{eqnarray}
\Delta (x_{\perp })=-L(x_{\perp })\theta _{51}\theta _{52}\ scalar
\end{eqnarray}
\begin{eqnarray}
\Delta (x_{\perp })=J(x_{\perp })\hat{P}\cdot \theta _1\hat{P}\cdot \theta
_2\ time\ like\ vector
\end{eqnarray}
\begin{eqnarray}
\Delta (x_{\perp })={\cal G}(x_{\perp })\theta _{1\perp }\cdot \theta
_{2\perp }\ space\ like\ vector
\end{eqnarray}
\begin{eqnarray}
\Delta (x_{\perp })={\cal F}(x_{\perp })\theta _{1\perp }\cdot \theta
_{2\perp }\theta _{51}\theta _{52}\hat{P}\cdot \theta _1\hat{P}\cdot \theta
_2\ tensor\ (polar).
\end{eqnarray}
The constraint equations for QED presented at the beginning of this paper
are generated in this form by taking $L=0={\cal F}$ and ${\cal G}=-J=-{\frac 
12}ln(1-2{\cal A}/w)$. For the QCD calculations which we performed above $L$
is non zero and determined \cite{sclfrm} by the $S$ of Eq.(\ref{sclr}) while
the vector interaction is written in terms of ${\cal A}$ of Eq.(\ref{vctr})
and just as for QED has the Feynman gauge combination $J=-{\cal G}$. For the
axial counterparts the hyperbolic forms of our constraints are (note the
minus sign) 
\end{mathletters}
\begin{mathletters}

\begin{eqnarray}
{\cal S}_1\psi =(ch(\Delta ){\bf S}_1-sh(\Delta ){\bf S}_2)\psi =0
\end{eqnarray}
\begin{eqnarray}
{\cal S}_2\psi =(ch(\Delta ){\bf S}_2-sh(\Delta ){\bf S}_1)\psi =0,
\end{eqnarray}
in which ${\bf S}_1$ and ${\bf S}_2$ are defined as before while the
interactions appear through 
\end{mathletters}
\begin{mathletters}

\begin{eqnarray}
\Delta (x_{\perp })=C(x_{\perp })/2\ pseudoscalar
\end{eqnarray}
\begin{eqnarray}
\Delta (x_{\perp })=H(x_{\perp })\hat{P}\cdot \theta _1\hat{P}\cdot \theta
_2\theta _{51}\theta _{52}\ time\ like\ pseudovector
\end{eqnarray}
\begin{eqnarray}
\Delta (x_{\perp })=I(x_{\perp })\theta _{1\perp }\cdot \theta _{2\perp
}\theta _{51}\theta _{52}\ space\ like\ pseudovector
\end{eqnarray}
\begin{eqnarray}
\Delta (x_{\perp })=Y(x_{\perp })\theta _{1\perp }\cdot \theta _{2\perp }%
\hat{P}\cdot \theta _1\hat{P}\cdot \theta _2\ tensor\ (axial).
\end{eqnarray}
Future research will determine the relative importance of the interactions
other than scalar and vector in meson spectroscopy.

We conclude by listing important features realized by these two body Dirac
equations which are incompletely realized if present at all in other
approaches to the relativistic two-body problem. A) manifest covariance, B)
simple three-dimensional Schr\"{o}dinger-like forms similar to their
nonrelativistic counterparts, C) spin dependence dictated by the Dirac-like
structure of the equations (not put in by hand), D) thoroughly tested in
QED: the equations reproduce the correct perturbative results when solved
nonperturbatively (they qualify as bona fide wave equations less likely to
produce spurious nonpeturbative effects), E) the close connection to the one
body Dirac equations automatically eliminates the ad hoc introduction of
cutoff parameters, G) its potentials are determined through straightforward
connection to perturbative quantum field theory or introduced
semiphenomenologically.

\end{mathletters}


\begin{references}
\bibitem{sclfrm}  The relation between $S$ and $L$ (and ${\cal G}$) is given
in \cite{cra84,cra87,cra88,cra94}. One finds $M_i^2=m_i^2+{\rm e}^{2{\cal G}%
}(2m_1m_2S/w+S^2)$.
\end{references}
\end{document}